\documentclass[10pt]{article}

\usepackage{graphicx}
\usepackage{amssymb}
\usepackage{amsmath}

\usepackage{hyperref}
\usepackage{bbm}

\usepackage[latin1]{inputenc}
\usepackage{epsfig}
\usepackage{subfigure}

\usepackage{color}

\usepackage{epstopdf}
\newcommand{\argmax}{\operatornamewithlimits{argmax}}

\def\E{{\mathbb E}}
\def\sumi{\sum_{i=1}^n} 
\def\sumj{\sum_{j=1}^J} 
\def\sumk{\sum_{k=1}^K}

\def\etal{{\em et al. }}

\usepackage[total={17cm, 23.5cm},
top=3.5 cm, left=2.5 cm, right = 2.5 cm]{geometry}
\usepackage{authblk}

\title{Multi-resource fairness: Objectives, algorithms and performance}
\author[1]{Thomas Bonald}
\author[2]{James Roberts\thanks{The authors are members of the LINCS, Paris, France. See www.lincs.fr.}}
\affil[1]{T\'el\'ecom ParisTech, Paris, France}
\affil[2]{IRT System-X, Paris-Saclay, France}

\begin{document}

\maketitle

\begin{abstract} 
Designing efficient and fair algorithms for sharing multiple resources between heterogeneous demands is becoming increasingly important. Applications include compute clusters shared by multi-task jobs and routers equipped with middleboxes shared by flows of different types. We show that the currently preferred objective of Dominant Resource Fairness has a significantly less favorable efficiency-fairness tradeoff than alternatives like Proportional Fairness and our proposal, Bottleneck Max Fairness. In addition to other desirable properties, these objectives are equally strategyproof in any realistic scenario with dynamic demand.
%In particular, since demand is in fact \emph{dynamic}, the question of identifying allocations that are strategyproof for a \emph{fixed} set of transactions becomes a moot point.

 \end{abstract}

\section{Introduction}
Multi-resource fairness has recently received a lot of attention thanks mainly to two papers by A. Ghodsi and colleagues from UC Berkeley \cite{Ghodsi2011, Ghodsi2012}. In the present paper we argue that the objective of dominant resource fairness (DRF), advocated by Ghodsi \etal in \cite{Ghodsi2011} and \cite{Ghodsi2012} for resource sharing in compute clusters and routers equipped with middleboxes, respectively, is misguided. More classical sharing objectives, like proportional fairness (PF), achieve a better efficiency-fairness tradeoff. Such objectives were discarded in the UC Berkeley work as being vulnerable to manipulation by users seeking to gain more than their fair share by falsely stating their requirements. However, we argue that the advocated\emph{ strategyproofness} property, possessed by DRF but not PF in a context of \emph{static} demand, is not in fact discriminating when considering the more realistic context of \emph{dynamic} demand. 

We have already made the case  in a recent paper for proportional fair sharing of compute cluster resources \cite{BR2014}%\footnote{This paper by the present authors is to be published. An anonymized copy is available here: \href{https://files.acrobat.com/a/preview/2962031d- 6a3f-4456-af85-1c70b59e2747}{https://files.acrobat.com/a/preview/2962031d- 6a3f-4456-af85-1c70b59e2747}}
. Here we extend those arguments to the case of router resources and introduce a new allocation objective called bottleneck max fairness (BMF). We propose packet-based algorithms to realize PF and BMF that have comparable complexity to the DRFQ algorithms proposed to realize DRF \cite{Ghodsi2012}.

Multi-resource sharing in a compute cluster consists in launching appropriate numbers of tasks of multi-task jobs. Each task of a given job has its particular requirements for CPU, RAM and other resources. In practice, various constraints on the placement of tasks on physical machines need to be taken into account but, for present purposes, we follow Ghodsi \etal \cite{Ghodsi2011} and assume resources of the same type are assembled in homogeneous pools. %We also assume all tasks of the same job have the same requirements. 
The question is, how should a central scheduler determine the numbers of simultaneous tasks to run for all the currently active jobs, ensuring efficiency and some measure of fairness. 

Routers increasingly employ middleboxes to process packets and these constitute potential bottlenecks for flows, in addition to link bandwidth. Different flows have different per-packet resource requirements (e.g., some require complex packet processing, others none) and the issue here is what packet rates should be imposed for concurrent flows in order to efficiently and fairly share all types of resource. 

Bandwidth sharing in a network is a particular form of multi-resource sharing. The problem is generally simpler since all flows sharing a given link are assumed to have identical requirements. This assumption is not true, however, for some wireless links where the resource to be allocated is spectrum time and the amount required to realize a given bit rate varies considerably between terminals, depending on their particular radio conditions. It is important to share wireless and wired links in a manner that adequately balances efficiency and fairness.

We first describe DRF, PF and BMF sharing objectives in abstract terms common to the above three application contexts. We define the corresponding allocations and discuss their significant characteristics.  Algorithms are then presented for realizing the respective allocations. Our main original contribution here is to identify practical, packet-based algorithms that realize PF and BMF sharing between flows in a router or network.  The following  section on performance presents numerical results that justify our claim that DRF is not in fact the best objective. Under dynamic demand, where jobs or flows arrive over time and have finite size, completion times are consistently and significantly smaller with either PF or BMF allocations.

%
%this position paper = sharing algorithm performance should be measured in terms of completion times. our analysis suggests DRF not best, prefer PF or close (from network bandwidth sharing).

%sharing notions - jobs/ tasks v flows/packets, same fluid analogy, different algorithms. In a recent paper we showed sharing objectives like PF preferred to DRF for cluster. Here, mainly consider objectives and algorithms for flow/packet case

%static population of jobs - economics approach: strategyproofness, utility max - ``best papers'' at Sigcomm 2012, Infocom 2012

%but completion times depend on dynamics - strategyproofness not relevant if any user cannot know competing flow characteristics

%performance in dynamic traffic (analogy to show dynamics need to be taken into account? road traffic? telephone traffic? internet performance? )

%assume uniform resource pools, ignoring constraints on placement 

%contributions: refer to \cite{BR2014}  for cluster application, main focus here on network applications, fairness objectives: DRF, PF, BMF; algorithms: most deprived job, FQ algorithms + simple transport; performance: under dynamic demand.

\section{Objectives}
We define DRF, PF and BMF multi-resource sharing objectives with respect to a fluid model where resources are assumed infinitely divisible. The model is abstract and, rather than jobs or flows, we refer to \emph{transactions}. These are assumed divisible into a large number of infinitesimal components having homogeneous resource usage characteristics. 

\subsection{A fluid model}

%examples (?): note task is particle of job, requires fraction of resources for given time, particle of flow is packet, requires processing (compute or transmission). Notions of simultaneous resource use (particles = tasks) and successive resource use (particles = packets) 

%\begin{tabular}{l l l l }
%  \hline   
%  & cluster & CDMA & middlebox \\   
%    \hline  
% transaction & job & flow & flow \\
% resource & CPU & time slots/s & proc. time \\                 
% capacity & cores & (slot length)$^{-1}$ & GHz\\
% particle & task & 1 bit/s & 1 packet/s \\
%  \hline  
%\end{tabular}

%
%generally assume particles identical with respect to resource requirements (though this can be relaxed for practical implementation, eg, variable packet size, jobs with heterogeneous tasks) - for packets, assume sequential resource use (not parallel or fixed window implementations don't apply (?))

Consider $J$ infinitely divisible resources of normalized capacity 1 to be shared by  $n$ transactions indexed by $i$. Each transaction $i$ requires amounts of each resource in fixed proportions and the amount allocated determines the rate at which the transaction progresses. We denote the requirements of transaction $i$ by a vector $a_i= (a_{i1},\ldots,  a_{iJ})$ where, for definiteness, and without loss of generality, we set $\max_j (a_{ij})$ = 1. A resource for which $a_{ij}=1$ is called a \emph{dominant resource} for transaction $i$. 

An allocation is defined by a vector of real numbers $\varphi=(\varphi_1,\ldots,\varphi_n)$ such that $\varphi_i a_{ij}$ is the fraction of resource $j$ allocated to transaction $i$.  The allocation must satisfy capacity constraints:
\begin{equation}\label{eq:capa}
\sum_{i=1}^n \varphi_i a_{ij} \le 1, \;\; \mathrm{ for }\; j=1,\ldots,J.
\end{equation}

Following Ghodsi \etal \cite{Ghodsi2011}, allocations should have the following properties. They should be \emph{Pareto efficient} in that no fraction of any resource should be left needlessly idle. They should offer a \emph{sharing-incentive} in that any transaction is assured at least a fraction $1/n$ of its dominant resource. They should be \emph{single resource fair} in the sense that $\varphi_i a_{i1} = 1/n$ when $J=1$. It may readily be verified that the following three allocations have these properties \cite{Ghodsi2011,BR2014,Dolev2012}.

\subsection{Dominant resource fairness}
\label{sec:drfobj}

Dominant resource fairness (DRF) is the unique Pareto efficient allocation where transactions obtain fractions of their dominant resource that are as equal as possible. Equality may not be possible if some $a_{ij}$ are zero \cite{Parkes2012}. The main advantage of DRF identified in \cite{Ghodsi2011} is that, in addition to the above-mentioned properties, it is \emph{strategyproof}. This means a transaction cannot gain a greater share of resources by boosting some component of its requirement vector. %It is shown in \cite{Ghodsi2011} that this property in not shared by other allocations where a transaction share can be increased to the detriment of others by judiciously increasing the declared requirement of some non-dominant resource(s).

\subsection{Proportional fairness}
\label{pfobj}

An allocation $\varphi$ is proportional fair (PF) if, for any other allocation $\phi$ satisfying \eqref{eq:capa}, the sum of proportional changes is negative or zero, $\sum_i (\phi_i - \varphi_i)/\varphi_i \le 0$ \cite{Kelly1998}. Equivalently, $\varphi$ maximizes  $\sum_i \log\varphi_i$ subject to \eqref{eq:capa} and can thus be said to maximize social welfare assuming a logarithmic utility function. 
For a single resource, the allocation is such that the $\varphi_i a_{i1}$ are equal, fulfilling single resource fairness.  It can be shown that PF is the only utility-based allocation with this property \cite{BR2014}. The PF allocation is unique.

While PF was introduced in \cite{Kelly1998} as an objective for sharing bandwidth in a wired network, it was advocated independently by Tse and co-authors as the preferred allocation for a time-shared wireless downlink channel \cite{Viswanath2002}. Here $1/a_{i1}$ represents the number of bits transmissible per constant length time slot and depends on the radio conditions of receiver $i$. Allocations $\varphi_i$ are measured in bit/s and the PF allocation is such that each transaction receives the same slot rate $\varphi_i a_{i1}$.

%PF is realized in particular by wireless networks like HSDPA that assign equal numbers of constant length time slots to ongoing flows. The requirements $a_{i1}$ relate to the flow bit rate per slot which depends on the radio conditions experienced by the corresponding terminal \cite{Tse?}. Proportional fairness is also advocated as an objective for bandwidth sharing in wired networks since it arguably realizes greater social welfare than max-min \cite{KMT98}.

PF is \emph{not} strategyproof: if a user knows the requirements of concurrent transactions, it can obtain a bigger allocation by judiciously increasing its requirement for some non-dominant resource. For example, consider $n=2$ jobs and $J=2$ resources with $a_1=( 1/2, 1) $ and $a_2=( 1, 1/2) $. The PF allocation is then $\varphi=(2/3, 2/3)$. If transaction 1 claims $2/3$  of resource 1 instead of $1/2$, we find $\varphi'=(3/4, 1/2)$ yielding a bigger share for transaction 1 at the expense of transaction 2.

Note, however, that if a user does not know the other requirements, an ill-chosen modification can instead lead to it receiving a smaller share. Continuing the above example, if transaction 1 claims $a_1=(1,1)$, the PF allocation would be $\varphi'=(1/2, 1/2)$. Both transactions lose out compared to the result of the truthful declaration, $a_1=(1/2,1)$. 

%- strategy proofness but note that strategy profitable in static traffic (eg, user cheats, slows concurrent jobs, still there when same user launches a new job)
\subsection{Bottleneck max fairness}
Dolev \etal proposed the ``no justified complaints'' objective as an alternative to DRF  \cite{Dolev2012}. To be concise, we adopt a simplified definition of this objective: a transaction has no justified complaint if it receives at least a fraction $1/n$ of at least one fully allocated resource. The latter constitutes a bottleneck and Pareto efficient allocations fulfilling this objective for all users have been said to realize ``bottleneck-based fairness'' \cite{Zeldes2013, Gutman2012}. 

Unlike DRF and PF, there is in general no unique bottleneck-based fair allocation according to the above definition. We restrict the class of such allocations somewhat by requiring in addition that every transaction $i$ receives an allocation $\varphi_i a_{ij}$ of some bottleneck resource $j$ that is \emph{maximal} for that resource. We call such allocations ``bottleneck max fair'' (BMF). %Relevance is that this allocation can be easily achieved for packet-based implementation (see below).

It can be shown that a system with 2 resources has a unique BMF allocation but that, for 3 or more resources, there can be a continuum of allocations realizing the BMF sharing objective. For example, consider a 3 resource system with 3 transactions having requirement vectors  (1, 1, 1), (1, 1/2, 3/4) and (1/2, 1, 3/4). Allocations $\varphi^{(0)} = (2/5, 2/5, 2/5)$ and $\varphi^{(1)} = (1/3, 4/9, 4/9)$ both satisfy the definition. In fact, allocations  $\varphi^{(x)} = (2/5-x/15, 2/5 + 2x/45, 2/5 + 2x/45) $ for $0 \le x \le 1$ are all BMF. %Note that all these allocations may be considered to be ``fair'' in any intuitive sense. 
%The allocation $\varphi^{(0)}$ is max-min BMF with equal maximal bottleneck shares of 2/5 for each transaction.In the latter case, there is clearly one ``max-min BMF'' allocation that maximizes the minimum maximum bottleneck share over all transactions.  

It can be shown that BMF coincides with PF for $n=2$ and $J=2$ and is therefore not strategyproof.

\section{Algorithms}
\label{sec:algorithms}

We discuss algorithms to realize DRF, PF and BMF, successively for the fluid model, for a compute cluster shared by multi-task jobs and for router or network resources shared by flows of packets.
 
\subsection{Dominant resource fairness}
\label{sec:drfalg}
To realize the ideal DRF allocation of Section \ref{sec:drfobj} one can employ a water-filling algorithm \cite{Ghodsi2011, Parkes2012}: with the dominant resource requirement normalized to 1, the $\varphi_i$ are increased at the same rate until some resource is fully used; transactions using that resource are frozen while rates of the others with non-zero requirements on non-saturated resources are increased together until a second resource is full; the process continues until all the $\varphi_i$ are frozen.

Ghodsi \etal \cite{Ghodsi2011} also show how the DRF allocation can be realized in practice for compute cluster resources. As resources are freed on task completion they are re-allocated preferentially to the most deprived job. This is the job for which the difference between the ideal and current allocations of its dominant resource is greatest.

To share router resources between flows, we consider the simpler case where all packets of the same flow have the same requirement vector. The  memoryless DRFQ algorithm defined by Ghodsi \etal \cite{Ghodsi2012} then applies. 
We suppose there is a fixed-size window of $W$ packets for each flow and that buffers are sized for no loss. This window might be realized within a router by creating an ingress queue and only admitting packet $k$ when packet $k-W$ has finished processing at all resources. 

Memoryless DRFQ determines the order in which packets are served at each resource through an adaptation of start-time fair queuing (SFQ) \cite{Goyal1997}. The virtual start time $S^k_i$ of packet $k$ of flow $i$ is determined recursively,
\begin{equation}
S^k_i = \max \left(V(u^k_i),S^{k-1}_i+\max_j\{a_{ij}\}\right),  \label{eq:stdrf}
\end{equation}
where $u^k_i $ is the packet arrival time and the virtual time function of real time $t$, $V(t)$, is set equal to the largest start time at $t$ of any packet to have begun service at any resource.  Packets are served at each resource in increasing order of the $S^k_i$.  Note that, with our normalized requirements, the max in \eqref{eq:stdrf} is identically equal to 1. %(different to \cite{Ghodsi2012} but there is spurious notation).

This algorithm is generalized in \cite{Ghodsi2012}, as ``dovetailing DRFQ'', to account for successive packets of the same flow having different requirements. 
 
%Note a problem if only one flow active and first resource processing time smaller than subsequent processing time: packets accumulate at second resource queue;  $S(p^k_i)$ increases with $k$ meaning virtual time also; new flow must wait for admitted flow $i$ packets to clear. In practice, cannot accept packets indefinitely since buffers finite. Losses occur, interaction with congestion control.

%Suppose fixed window size to represent congestion control: more exactly, buffer at router ingress  limits maximum number of packets per flow, one leaves, another is allowed in; buffers sized to avoid loss.

 \subsection{Proportional fairness}
 \label{sec:pfalg}

The PF allocation for the fluid model can be derived on applying the Karuch-Kuhn-Tucker theorem to maximize $\sum \log(\varphi_i)$ subject to capacity constraints \eqref{eq:capa}. We must find Lagrange multipliers $\nu_j$ satisfying
\begin{equation}\label{eq:kkt}
{1\over \varphi_i}=\sumj a_{ij}\nu_j,
\end{equation}
for $i=1,\ldots,n$, where $\nu_j \ge 0$ and $\nu_j> 0$ if and only if $\sumi \varphi_i a_{ij}=1$.
%This optimization is not hard since the number of resource pools is typically small. 

The optimal solution can be used in a practical task-based algorithm for sharing a compute cluster. This consists in applying the same ``most deprived job'' approach from Section \ref{sec:drfalg} with, of course, an alternative criterion to measure deprivation \cite{BR2014}.

An algorithm for packet-based PF can be derived on adapting the analysis of Massouli\'e and Roberts for network bandwidth sharing \cite{Massoulie2002}. We denote the number of flow $i$ packets waiting or in service at resource $j$ by $Q_{ij}$ and let $Q(j)=\sum_i Q_{ij}$. Assuming as above that each flow maintains a fixed window of $W$ packets, we have the conservation equation,
\begin{equation}
W =  \sum_j Q_{ij}.
\label{eq:conserve2}
\end{equation}
We assume the system attains a stable regime where the $Q_{ij}$ are positive constants when resource $j$ is a bottleneck, or zero otherwise. It turns out that serving packets at rates proportional to  $Q_{ij}/a_{ij}$ yields the PF allocation. We have,
$$\sum_i \varphi_i a_{ij} = 1 \Rightarrow {Q_{i'j} \over Q_{ij}} = {\varphi_{i'} a_{i'j} \over \varphi_{i} a_{ij}}$$
and summing over $i'$ yields $Q_{ij}=\varphi_i a_{ij}  Q(j)$.  On substituting for $Q_{ij}$ in \eqref{eq:conserve2} we derive,
$$ {1\over \varphi_i} = \sum_j a_{ij} {Q(j)\over W} $$
where $Q(j)\ge 0$ and $Q(j)>0$ if and only if $\sum_i \varphi_i a_{ij}=1$.
Comparison with \eqref{eq:kkt} shows that the $Q(j)/W$ coincide with the Lagrange multipliers $\nu_j$ and, therefore, that $\varphi$ is indeed the unique PF allocation.

To realize the required packet rates we can adapt the SFQ algorithm. Start times $S^k_{ij}$ are defined recursively and independently for each resource $j$,
\begin{equation}
S^k_{ij} = \max \left(V_j(u^k_{ij}),S^{k-1}_{ij}+a_{ij}/Q_{ij}\right) ,  \label{eq:stpf}
\end{equation}
where $u^k_{ij}$ is the packet arrival time at resource $j$ and virtual time $V_j(t)$ is the start time at $t$ of the last packet to have begun service at $j$.

This algorithm can be applied when the $a_{ij}$ vary from packet to packet realizing a dovetailing PF. %, as in dovetailing DRFQ, though the precise properties of the resulting allocation remain to be evaluated. 

 \subsection{Bottleneck max fairness}
  \label{sec:bmfalg}
For small systems, it is possible to determine a BMF allocation by testing the feasibility of all possible one-to-one mappings of transactions to potential bottlenecks. For a mapping to be feasible, it must be possible to find values $\varphi_i$ satisfying \eqref{eq:capa} such that every resource $j$ with at least one mapped transaction is a bottleneck ($\sum \varphi_i a_{ij}=1$) and the allocations of transactions mapped to $j$ are equal (i.e., $\varphi_i a_{ij} =  \varphi_i' a_{i'j}$ if $i$ and $i'$ are mapped to $j$). 

For a system limited to two resources, it can be shown that this procedure yields the unique BMF allocation. However, to prove the existence of a BFM allocation in general is challenging (see \cite{Dolev2012} and  \cite{Gutman2012}) though the following practical algorithms suggest this is true. 

The most deprived job approach can be adapted to realize BMF in a cluster. For every job $i$ we note its rank at each resource $j$: jobs with the biggest allocation have rank $1$, jobs of rank $k$ for $k>1$ have a smaller share than $k-1$ other jobs. The deprivation status of job $i$ is its minimum rank on a bottleneck resource. Tasks are launched preferentially for the job whose current status is largest.

For the router application, BMF can be realized by imposing local weighted max-min fairness at each resource with respective weights $1/a_{ij}$. To demonstrate this, we again assume the existence of a steady state and adapt arguments from \cite{Massoulie2002}. We assume each flow $i$ maintains a window of $W$ packets and, to include the possibility of remotely located resources (as in the network application), we introduce a round trip propagation time $T_i$.  The following conservation relations generalize \eqref{eq:conserve2},
\begin{equation}
W = \varphi_i T_i + \sum_j Q_{ij}.
\label{eq:conserve}
\end{equation}
Relation \eqref{eq:conserve} shows that for every transaction $i$, as long as  $W>\varphi_i T_i$,  there is at least one bottleneck resource $j$ such that $Q_{ij}>0$ and $\sum_i \varphi_i a_{ij}=1$. Moreover, the weighted fair queuing scheduler at $j$ ensures $\varphi_i a_{ij} = \max_{i'} (\varphi_{i'} a_{i'j})$, completing the BMF defining conditions.

To realize BMF using SFQ, start times must be calculated as follows,
\begin{eqnarray}
S^k_{ij} = \max \left(V_j(u^k_i),S^{k-1}_{ij}+a_{ij}\right) ,  \label{eq:stbmf}
\end{eqnarray}
where $u^k_i$ and $V_j(t)$ are as defined in Section \ref{sec:pfalg}.

As for DRF and PF, this algorithm does not need the $a_{ij}$ of successive packets to be constant, leading to a corresponding dovetailing algorithm.  
 
\section{Performance}
While the sharing algorithms are defined with respect to a fixed set of transactions, their performance can only be realistically appraised under dynamic demand where transactions occur over time, each bringing a finite amount of work to be accomplished. We propose a simple Markovian demand model and use it to compare the completion time performance of DRF, PF and BMF.

\subsection{Markovian demand model}
We suppose transactions belong to one of $K$ classes and transactions of class $k$ arrive as a Poisson process of rate $\lambda_k$. Class $k$ transactions have requirement vector $(a_{k1},\ldots,a_{kJ})$ and bring an exponentially distributed amount of  work (number of tasks $\times$ mean task duration or size of a flow in bytes, say) of mean $1/\mu_k$. The state vector $(n_1,\ldots,n_K)$, giving the current number of transactions in progress, is then a Markov process with component-$k$ birth rate $\lambda_k$ and death rate $n_k\varphi_k \mu_k$. 

This process is stable as long as loads $\rho_k =\lambda_k/ \mu_k$ satisfy the following inequalities,
\begin{equation}
\sumk \rho_k a_{kj}<1,
\label{eq:capacity}
\end{equation}
for $j=1,\ldots,J$. In this case, the process has a stationary distribution $\pi(n)$ from which we can compute performance measures like expected completion times. 

In the following we compare algorithm performance via the mean service rate $\gamma_k$, defined as the ratio of the mean work $1/\mu_k$ to the mean completion time. Its reciprocal is thus a normalized completion time. Using Little's law, we find $\gamma_k=\lambda_k/\E(n_k)$ where $\E(n_k)$ is the mean number of class $k$ transactions in progress.
See \cite{Bonald2006} for more information on this model including a discussion on the non-criticality of Poisson and exponential assumptions.

%
%Markov traffic model - simple Montecarlo simulations - insensitivity implies results valid for more realistic traffic (cf \cite{Bonald2006})

%$\rho_i$ = demand of class $i$

%resource load proportional to $\sum_i \rho_i a_{ij}$

%performance expressed in terms of service rate  $\gamma_i$ ...

\subsection{Strategyproofness}
First reconsider the notion of strategyproofness in the context of dynamic demand. While it may be possible for an agent to know the requirement vectors of competing classes, it is hardly reasonable to suppose knowledge of the current numbers of transactions in progress. Moreover, these numbers change rapidly and a good strategy in one state may rapidly become disadvantageous as the state  changes.

For instance, PF was shown in Section \ref{pfobj} to be vulnerable to a falsely declared requirement vector: transaction 1 with requirement vector $a_1= (1/2,1)$ competing with transaction 2 with vector $(1,1/2)$ was shown to gain from a falsely declared vector $(2/3,1)$. If now, there are 2 transactions like the second, the same strategy results in transaction 1 receiving $\varphi_1=1/2$, less than the 2/3 obtained with a true declaration.

We have identified no winning strategy for any of the considered allocations and, in the absence of any counter-examples, consider them all to be equally strategyproof. Note, in particular, that an increase in any requirement $a_{ij}$ can diminish the system capacity region \eqref{eq:capacity} degrading performance for all.

\subsection{Performance of the fluid model}
\label{sec:fluidperf}

We illustrate the relative performance of DRF, PF and BMF using a numerical example. Two resources are shared by 3 classes of transaction with requirement vectors $a_1=(.1,1)$, $a_2=(1,.1)$, $a_3=(1,1)$. The figures plot realized service rates $\gamma_k$ against the load of the heaviest loaded resource ($\argmax_j (\sum_i \rho_i a_{ij})$).

\begin{figure}[t]
 \centering
\subfigure[BMF v DRF]{ \includegraphics[scale=.65]{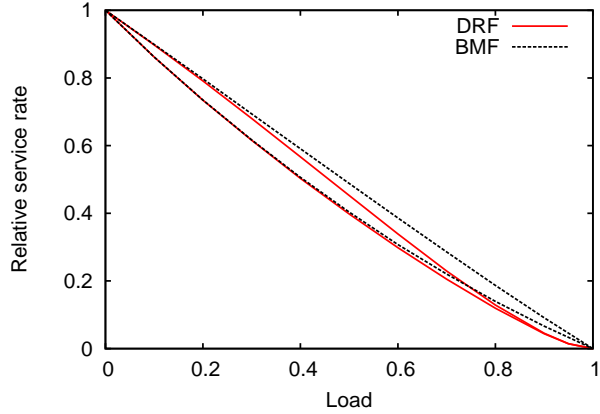}}
\subfigure[BMF v PF]{ \includegraphics[scale=.65]{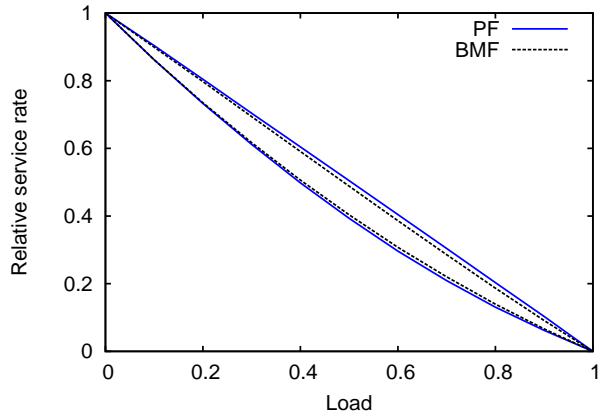}}
 \caption{Service rates $\gamma_k$ against resource load for BMF and DRF with balanced load: $a_1 = (.1,1)$, $a_2= (1,.1)$, $a_3=(1,1)$; $\rho_1=\rho_2=\rho_3$;  for each allocation we have, $\gamma_1=\gamma_2>\gamma_3$.}
 \label{fig:fluid-bal}
\end{figure}

Figure \ref{fig:fluid-bal} corresponds to balanced load  $\rho_1=\rho_2=\rho_3$ and shows performance is similar for all three allocations. BMF is very close to PF while both are somewhat better than DRF, especially at high load.  

\begin{figure}[t]
 \centering
\subfigure[BMF v DRF]{ \includegraphics[scale=.65]{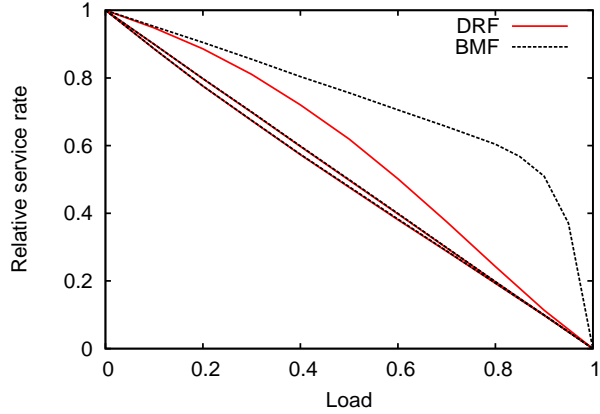}}
\subfigure[BMF v PF]{\includegraphics[scale=.65]{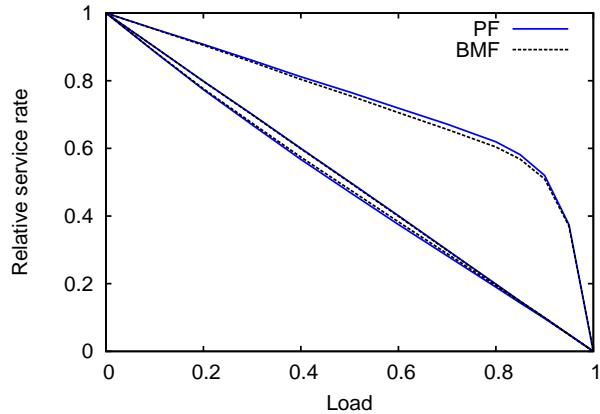}}
\caption{Service rates  $\gamma_k$ against resource 2 load for BMF and DRF with unbalanced load: $a_1 = (.1,1)$, $a_2= (1,.1)$, $a_3=(1,1)$; $\rho_1=4\rho_2=4\rho_3$; for each allocation we have, $\gamma_2>\gamma_1>\gamma_3$.}
 \label{fig:fluid-unbal}
\end{figure}

The difference in performance is accentuated under unbalanced load, as illustrated in Figure \ref{fig:fluid-unbal} for the case $\rho_1=4\rho_2=4\rho_3$. In this scenario resource 1 has less than half the load of resource 2. The plots show that strict fairness imposed by DRF prevents class 2 transactions from fully exploiting this. The service rates of PF and BMF are roughly equivalent, both yielding a better efficiency-fairness tradeoff with a significant gain in $\gamma_2$ for negligible reductions in $\gamma_1$ and $\gamma_3$.

%\begin{figure}[t]
% \centering
% \includegraphics[scale=.5]{fluid-bal-bmf-pf.eps}
% \caption{Service rates  $\gamma_k$ for BMF and PF for balanced load: $a_1 = (.1,1)$, $a_2= (1,.1)$, $a_3=(1,1)$; $\rho_1=\rho_2=\rho_3$; for each allocation, $\gamma_1=\gamma_2>\gamma_3$.}
% \label{fig:fluid-bal-bmf-pf}
%\end{figure}

%\begin{figure}[t]
% \centering
% \includegraphics[scale=.5]{fluid-unbal-bmf-drf.eps}
% \caption{Service rates  $\gamma_k$ for BMF and DRF for unbalanced load: $a_1 = (.1,1)$, $a_2= (1,.1)$, $a_3=(1,1)$; $\rho_1=4\rho_2=4\rho_3$; for each allocation, $\gamma_1>\gamma_2>\gamma_3$.}
% \label{fig:fluid-unbal-bmf-drf}
%\end{figure}

%\begin{figure}[t]
% \centering
% \includegraphics[scale=.5]{fluid-unbal-bmf-pf.eps}
% \caption{Service rates  $\gamma_k$ for BMF and PF for unbalanced load: $a_1 = (.1,1)$, $a_2= (1,.1)$, $a_3=(1,1)$; $\rho_1=4\rho_2=4\rho_3$;  for each allocation, $\gamma_1>\gamma_2>\gamma_3$.}
% \label{fig: fluid-unbal-bmf-pf}
%\end{figure}

%
%\begin{figure}[t]
% \centering
% \includegraphics[scale=.3]{fx-balanced.pdf}
% \caption{Throughput v load for LPF, DRF and PF for balanced load.}
% \label{fig:compare}
%\end{figure}

%
%\begin{figure}[t]
% \centering
% \includegraphics[scale=.3]{fx-unbalanced.pdf}
% \caption{Throughput v load for LPF, DRF and PF for unbalanced load.}
% \label{fig:lpf}
%\end{figure}

%\begin{figure}[t]
% \centering
% \includegraphics[scale=.3]{fluid-balanced.pdf}
% \caption{Throughput v load for LPF, DRF and PF for balanced load.}
% \label{fig:compare}
%\end{figure}

\begin{figure*}[!ht]
\begin{center}
\subfigure[DRF]{ \includegraphics[scale=.65]{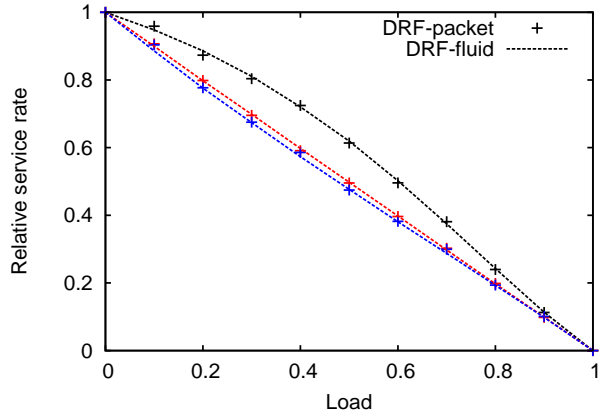}}
\subfigure[PF]{\includegraphics[scale=.65]{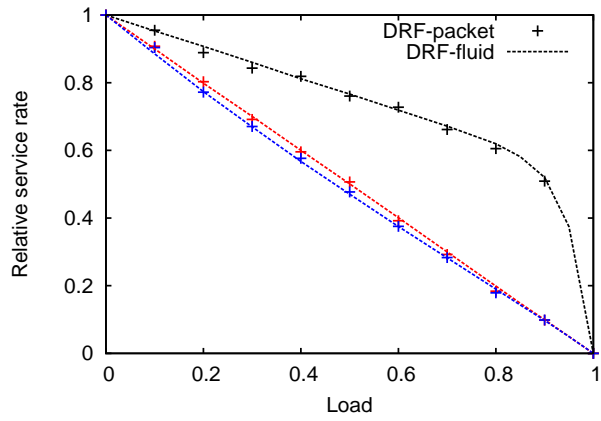}}
\subfigure[BMF]{\includegraphics[scale=.65]{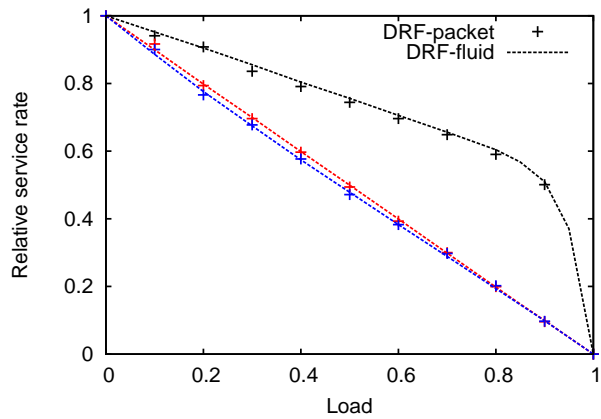}}
\caption{Service rates  $\gamma_k$ against resource 2 load for unbalanced load: $a_1 = (.1,1)$, $a_2= (1,.1)$, $a_3=(1,1)$; $\rho_1=4\rho_2=4\rho_3$; for each allocation, $\gamma_2>\gamma_1>\gamma_3$.}
\label{fig:pktfluid}
\end{center}
\end{figure*}

\subsection{Packet-based allocations}
It was shown in \cite{BR2014} that the serve-the-most-deprived-job algorithms yield service rates that retain the same comparative behaviour of DRF and PF illustrated in Section \ref{sec:fluidperf}. In this section we evaluate the effectiveness of the SFQ-based algorithms defined in Section \ref{sec:algorithms} for router resource sharing.
We assume the number of packets in each flow has a geometric distribution of mean 20000 and the window $W$ is set to 30 packets. These parameter choices do not critically impact the presented results. 

%\begin{figure*}[t!]
%    \centering
%    \begin{subfigure}[t]{0.3\textwidth}
%        \centering
%        \includegraphics[scale=.5]{pktflu-unbal-drf.eps}
%        \caption{DRF}
%    \end{subfigure}%
%    ~ 
%    \begin{subfigure}[t]{0.3\textwidth}
%        \centering
%        \includegraphics[scale=.5]{pktflu-unbal-pf.eps}
%        \caption{PF}
%    \end{subfigure}%
%    ~ 
%    \begin{subfigure}[t]{0.3\textwidth}
%        \centering
%        \includegraphics[scale=.5]{pktflu-unbal-bmf.eps}
%        \caption{BMF}
%    \end{subfigure}
%    \caption{Service rates  $\gamma_k$ against resource 2 load for unbalanced load: $a_1 = (.1,1)$, $a_2= (1,.1)$, $a_3=(1,1)$; $\rho_1=4\rho_2=4\rho_3$; for each allocation, $\gamma_1>\gamma_2>\gamma_3$.}
%\end{figure*}

Service rates are shown in Figure \ref{fig:pktfluid} for unbalanced load where crosses are results of  packet-based simulations for $10^5$ flow arrivals and lines are the fluid model results from the previous subsection. The results confirm that all three packet-based algorithms closely approximate the service rates of the ideal allocations. %Similar agreement was found for a number of other system configurations.

Given the similarity of the results for PF and BMF, it would be preferable to implement the algorithm for the latter as it is simpler in not requiring knowledge of per-flow queue lengths. Moreover, BMF is the only algorithm applicable when the propagation time is non-negligible, as in the case of a wireless access network.% with remotely located radio and backhaul resources. 

\section{Related work}
We limit the present discussion to the most relevant related work. 
DRF  \cite{Ghodsi2011} and ``no justified complaints''  \cite{Dolev2012} were placed in a more general economics framework in the work of Gutman and Nisan \cite{Gutman2012}. Joe-Wang \etal also generalized DRF by introducing two families of allocations that allow a controlled tradeoff between efficiency and fairness \cite{Wong2012}. All these objectives are evaluated assuming a fixed set of transactions. %We previously advocated the PF allocation \cite{BR2014} and introduce here BMF as a specialization of bottleneck-based fairness.

The ``serve the most deprived job'' approach introduced in \cite{Ghodsi2011} proves very versatile. It is used by Zeldes and Feitelson \cite{Zeldes2013} to implement bottleneck-based fairness and by Ghodsi and co-authors \cite{Ghodsi2013} to account for compatibility constraints in task placement. Our proposed implementations of PF and BMF for sharing cluster resources are further illustrations of this versatility.

The packet-based algorithms designed by Ghodsi \etal \cite{Ghodsi2012} to realize DRFQ for shared router resources are based on start-time fair queuing. Wang and co-authors have proposed alternative realizations that adapt DRR \cite{Shreedhar1995} to the multi-resource context \cite{Wang2013b}. Our implementations of PF and BMF rely on SFQ though it appears straightforward to substitute alternative fair queuing algorithms if required. 

The need to evaluate the performance of resource sharing objectives under dynamic demand is still not widely recognized. The paper by Massouli\'e and Roberts  \cite{Massoulie2000} was perhaps the first to note the importance of this while some of the most significant subsequent findings are summarized in Bonald \etal \cite{Bonald2006}. Our earlier paper \cite{BR2014} and the present work extend this analysis to the domain of multi-resource sharing.

\section{Conclusions}
Multi-resource sharing for efficiency and fairness is an old issue in networking with challenging new variants occurring in the domains of cluster computing and software routers. Recent prominent publications have led to the emergence of DRF and its apparent acceptance as the preferred sharing objective\footnote{DRF is implemented in the Hadoop Next Generation Fair Scheduler, for instance.}.

We have argued in this paper that this popularity is misplaced since alternative objectives like PF display a better efficiency-fairness tradeoff. This result is revealed on considering the completion time performance of alternative objectives under the realistic and crucial assumption that demand is dynamic: jobs and flows occur over time and their completion times depend critically on the implemented resource sharing objective. 

We have proposed BMF as a pragmatic alternative to PF. It has similar performance and can be realized more simply, especially when sharing router and network resources between flows.  It is clearly the most practical objective for a network integrating radio access and wired backhaul. Independent schedulers simply share their own resource equitably. BMF thus generalizes network-wide max-min fairness that is known to be realized by fair schedulers acting independently on each link \cite{Hahne91}.

Concerns about the vulnerability of objectives like PF and BMF to malicious gaming have been shown to be unfounded. While users can manipulate allocations by falsely boosting their declared requirement of non-dominant resources, their gain depends critically on knowing the requirements of competitors. Such knowledge is clearly inconceivable in the context of highly dynamic populations of active transactions occurring in a realistic model of demand.

The present work is clearly incomplete. It remains notably to more thoroughly evaluate the proposed algorithms under realistic demand models and accounting for practical resource usage constraints. From the theory point of view, it is necessary to carefully analyse the convergence of proposed algorithms and the stability characteristics of supposed equilibria.

% Present work incomplete: theoretical level, existence of steady state, convergence of algorithms to a steady state (or not). Practical implementation. more realistic transactions with heterogeneous component tasks/packets, evaluate BMF algorithm for cluster. Evaluation with trace data - demonstrate the truth of our claim for insensitivity.

\newpage

\bibliographystyle{abbrv}
\bibliography{multifair} 

\begin{thebibliography}{10}

\bibitem{Bonald2006}
T.~Bonald, L.~Massouli{\'e}, A.~Prouti\`{e}re, and J.~Virtamo.
\newblock A queueing analysis of max-min fairness, proportional fairness and
  balanced fairness.
\newblock {\em Queueing Syst. Theory Appl.}, 53(1-2):65--84, June 2006.

\bibitem{BR2014}
T.~Bonald and J.~Roberts.
\newblock Enhanced cluster computing performance through proportional fairness.
\newblock {\em Performance Evaluation}, 79:134--145, September 2014.

\bibitem{Dolev2012}
D.~Dolev, D.~G. Feitelson, J.~Y. Halpern, R.~Kupferman, and N.~Linial.
\newblock No justified complaints: On fair sharing of multiple resources.
\newblock In {\em Proceedings of the 3rd Innovations in Theoretical Computer
  Science Conference}, ITCS '12, pages 68--75, New York, NY, USA, 2012. ACM.

\bibitem{Ghodsi2012}
A.~Ghodsi, V.~Sekar, M.~Zaharia, and I.~Stoica.
\newblock Multi-resource fair queueing for packet processing.
\newblock In {\em Proceedings of ACM SIGCOMM 2012}, pages 1--12, New York, NY,
  USA, 2012. ACM.

\bibitem{Ghodsi2011}
A.~Ghodsi, M.~Zaharia, B.~Hindman, A.~Konwinski, S.~Shenker, and I.~Stoica.
\newblock Dominant resource fairness: Fair allocation of multiple resource
  types.
\newblock In {\em Proceedings of the 8th USENIX Conference on Networked Systems
  Design and Implementation}, NSDI'11, pages 24--24, Berkeley, CA, USA, 2011.
  USENIX Association.

\bibitem{Ghodsi2013}
A.~Ghodsi, M.~Zaharia, S.~Shenker, and I.~Stoica.
\newblock Choosy: Max-min fair sharing for datacenter jobs with constraints.
\newblock In {\em Proceedings of the 8th ACM European Conference on Computer
  Systems}, EuroSys '13, pages 365--378, New York, NY, USA, 2013. ACM.

\bibitem{Goyal1997}
P.~Goyal, H.~Vin, and H.~Cheng.
\newblock Start-time fair queueing: a scheduling algorithm for integrated
  services packet switching networks.
\newblock {\em Networking, IEEE/ACM Transactions on}, 5(5):690--704, Oct 1997.

\bibitem{Gutman2012}
A.~Gutman and N.~Nisan.
\newblock Fair allocation without trade.
\newblock In {\em Proceedings of the 11th International Conference on
  Autonomous Agents and Multiagent Systems - Volume 2}, AAMAS '12, pages
  719--728, Richland, SC, 2012. International Foundation for Autonomous Agents
  and Multiagent Systems.

\bibitem{Hahne91}
E.~Hahne.
\newblock Round-robin scheduling for max-min fairness in data networks.
\newblock {\em Selected Areas in Communications, IEEE Journal on},
  9(7):1024--1039, Sep 1991.

\bibitem{Wong2012}
C.~Joe-Wong, S.~Sen, T.~Lan, and M.~Chiang.
\newblock Multi-resource allocation: Fairness-efficiency tradeoffs in a
  unifying framework.
\newblock In {\em INFOCOM, 2012 Proceedings IEEE}, pages 1206--1214, 2012.

\bibitem{Kelly1998}
F.~P. Kelly, A.~K. Maulloo, and D.~K.~H. Tan.
\newblock Rate control for communication networks: Shadow prices, proportional
  fairness and stability.
\newblock {\em The Journal of the Operational Research Society}, 49(3):pp.
  237--252, 1998.

\bibitem{Massoulie2000}
L.~Massouli\'e and J.~Roberts.
\newblock Bandwidth sharing and admission control for elastic traffic.
\newblock {\em Telecommunication Systems}, 15(1-2):185--201, 2000.

\bibitem{Massoulie2002}
L.~Massouli{\'e} and J.~Roberts.
\newblock Bandwidth sharing: Objectives and algorithms.
\newblock {\em IEEE/ACM Trans. Netw.}, 10(3):320--328, June 2002.

\bibitem{Parkes2012}
D.~C. Parkes, A.~D. Procaccia, and N.~Shah.
\newblock Beyond dominant resource fairness: extensions, limitations, and
  indivisibilities.
\newblock In {\em ACM Conference on Electronic Commerce}, pages 808--825. ACM,
  2012.

\bibitem{Shreedhar1995}
M.~Shreedhar and G.~Varghese.
\newblock Efficient fair queueing using deficit round robin.
\newblock {\em SIGCOMM Comput. Commun. Rev.}, 25(4):231--242, Oct. 1995.

\bibitem{Viswanath2002}
P.~Viswanath, D.~Tse, and R.~Laroia.
\newblock Opportunistic beamforming using dumb antennas.
\newblock {\em Information Theory, IEEE Transactions on}, 48(6):1277--1294, Jun
  2002.

\bibitem{Wang2013b}
W.~Wang, B.~Li, and B.~Liang.
\newblock Multi-resource round robin: A low complexity packet scheduler with
  dominant resource fairness.
\newblock In {\em ICNP 2013}, 2013.

\bibitem{Zeldes2013}
Y.~Zeldes and D.~G. Feitelson.
\newblock On-line fair allocations based on bottlenecks and global priorities.
\newblock In {\em Proceedings of the 4th ACM/SPEC International Conference on
  Performance Engineering}, ICPE '13, pages 229--240, New York, NY, USA, 2013.
  ACM.

\end{thebibliography}

\end{document}